\documentclass[a4paper,11pt]{article}
\pdfoutput=1
\usepackage{jcappub}
\usepackage{bm}
\usepackage{soul}
\usepackage{latexsym}
\usepackage{dcolumn}
\usepackage{amsfonts,amssymb,amsmath}
\usepackage{graphicx,epsfig}
\usepackage{psfrag}
\usepackage{braket}
\usepackage{caption}
\usepackage{subcaption}
\usepackage{graphicx}
\usepackage{rotating}
\usepackage{hyperref}
\usepackage{tikz}
\usepackage{comment}
\usepackage{float}

\hypersetup{
	unicode=false,          
	pdftoolbar=true,        
	pdfmenubar=true,        
	pdffitwindow=false,     
	pdfstartview={FitH},    
	pdftitle={My title},    
	pdfauthor={Author},     
	pdfsubject={Subject},   
	pdfcreator={Creator},   
	pdfproducer={Producer}, 
	pdfkeywords={keyword1} {key2} {key3}, 
	pdfnewwindow=true,      
	colorlinks=true,       
	linkcolor=red,          
	citecolor=cyan,        
	filecolor=magenta,      
	urlcolor=green,           
	linktocpage=true
}

\makeatletter
\gdef\@fpheader{}
\makeatother

\begin{document}

\title{Evolving Dark Energy or Evolving Dark Matter?}

\author{Xingang Chen and Abraham Loeb}

\affiliation{Institute for Theory and Computation, Harvard-Smithsonian Center for Astrophysics, 60 Garden Street, Cambridge, MA 02138, USA}

\abstract
{We show that the latest empirical constraints on cosmology, from a combination of DESI, CMB and supernova data, can be accounted for if a small component of dark matter has an evolving and oscillating equation of state within $-1<w<1$. From a fundamental physics perspective, this interpretation is more appealing than an evolving phantom dark energy with $w<-1$, which violates the null energy condition.}

\maketitle
		
\setcounter{page}{1}		

\section{Introduction}

The standard model of cosmology has a dark energy component that is equivalent to the cosmological constant. 
Recently, however, DESI and supernovae measurements \cite{DESI:2025zgx} suggested an intriguing preference that the equation of state (EOS) of the dark energy in terms of the ratio between pressure and energy density, $w=P/\epsilon$, may be time-dependent. Even more surprisingly, there is evidence suggesting that the EOS of the dark energy had $w<-1$ at early cosmic time and only crossed to the regime $w>-1$ more recently.

Sectors with $w<-1$ violate the null energy condition and can lead to instabilities. Although not impossible to formulate, such models often become pathological when being UV completed and theorized beyond the phenomenological parameterization. (See e.g.~\cite{Moghtaderi:2025cns} for a recent discussion.)

An interesting question is whether the effect observed in DESI DR2 \cite{DESI:2025zgx} can be induced by theories that do not violate the null energy condition.
In this paper, we propose a novel possibility. In our models, the dark matter sector and the dark energy sector are still independent of each other
and the dark energy is a cosmological constant\footnote{Other interesting possibilities include introducing some special interactions between dark matter and dark energy \cite{Khoury:2025txd} and considering effects of negative spatial curvature \cite{Chen:2025mlf}.}; however, a fraction of dark matter has a time-dependent EOS, $w(t)$, varying or oscillating around $w=0$ but always satisfying $w>-1$. We show that such models could be misinterpreted as models with evolving dark energy (including dark energy with $w<-1$), if one insists, when fitting data, that dark matter is entirely made of the standard cold dark matter (CDM) with $w=0$. We explore a variety of possibilities with dark matter having a time-dependent EOS, for example, contributed by a small component of an exotic type of dark matter with an evolving and/or oscillating EOS around zero.

Although the report from the recent DESI DR2 publication is the main focus of this paper, the idea illustrated in this paper may be useful in more general contexts beyond this particular claim. 

The structure of the paper is as follows. In Sec.~\ref{Sec:DESI DR2}, we describe the properties of the cosmic expansion rate with different kinds of non-standard dark energy, in particular, the best-fit model of the evolving dark energy highlighted in DESI DR2. In Sec.~\ref{Sec:Evolving DM}, we demonstrate the effects of dark matter on cosmology with different evolving and/or oscillating EOS, and show how these may be confused as an evolving, especially phantom, dark energy if the dark matter sector is assumed to be CDM. An example of the underlying fundamental physics of such models is explored in Sec.~\ref{Sec:EDM example} and related constraints are mentioned in Sec.~\ref{Sec:Constraints}. Finally, Sec.~\ref{Sec:Conclusions} summarizes the main conclusions of this paper.

\section{Non-standard or evolving dark energy models}
\label{Sec:DESI DR2}

Let us first examine the effect of various non-standard dark energy models on the evolution of the Hubble parameter as a function of scale factor.

For a non-standard dark energy with pressure $P_{\rm DE}$, energy density $\epsilon_{\rm DE}$ and an EOS
\begin{align}
    w_{\rm DE} =\frac{P_{\rm DE}}{\epsilon_{\rm DE}} \approx w_0 + w_a (1-a) ~,
\end{align}
the fluid equation
\begin{align}
    \dot\epsilon_{\rm DE} + 3\frac{\dot a}{a} (1+w_{\rm DE}) \epsilon_{\rm DE} =0
\end{align}
implies that
\begin{align}
    \frac{\epsilon_{\rm DE}}{\epsilon_{\rm DE0}}
    = e^{-3w_a(1-a)} a^{-3(1+w_0+w_a)} ~.
\end{align}
Together with the CDM component and assuming a flat universe, they lead to the following Friedmann equation
\begin{align}
    \frac{H^2}{H_0^2} = \Omega_m a^{-3} + \Omega_{\rm DE} \, e^{-3w_a(1-a)} a^{-3(1+w_0+w_a)} ~,
\end{align}
where $H_0$ is the Hubble parameter today, $\Omega_{\rm component}$'s denote today's density parameters of different components and they should add up to one for a flat universe. We use ``matter" to include both cold dark matter (CDM) and baryonic matter.

\begin{figure}[H]
  \centering
  \includegraphics[width=0.9\textwidth]{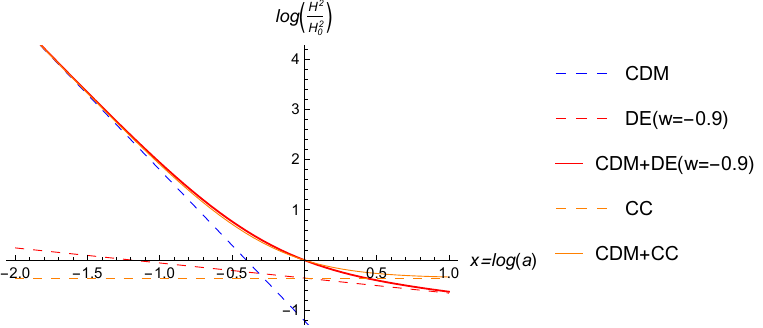}
  \caption{``CDM + DE ($w=-0.9$)" is an example with $\Omega_m=0.3$, $\Omega_{\rm DE}=1-\Omega_m$ and EOS of DE $w>-1$. ``CDM + CC" is the standard model with $\Omega_m=0.3$, $\Omega_{\rm DE}=1-\Omega_m$.
  The difference between ``CDM + DE ($w=-0.9$)" and ``CDM + CC" is small for $x\le 0$ in this figure and is amplified in Fig.~\ref{fig:DE_wLarger_DIFF}.
  In all figures, $\log(a)$ denotes the natural logarithm.
  }
  \label{fig:DE_wLarger}
  \vspace{1cm}
  \includegraphics[width=0.7\textwidth]{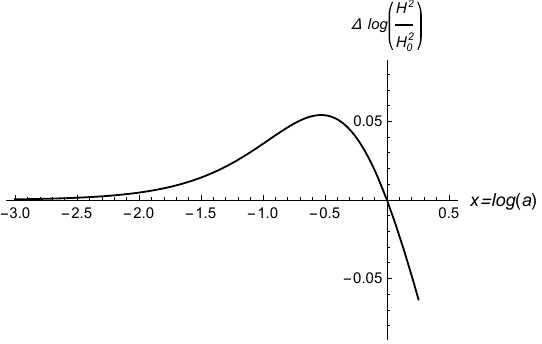}
  \caption{The difference between ``CDM + DE ($w=-0.9$)" and ``CDM + CC" in Fig.~\ref{fig:DE_wLarger}.}
  \label{fig:DE_wLarger_DIFF}
\end{figure}

\begin{figure}[H]
  \centering
  \includegraphics[width=0.9\textwidth]{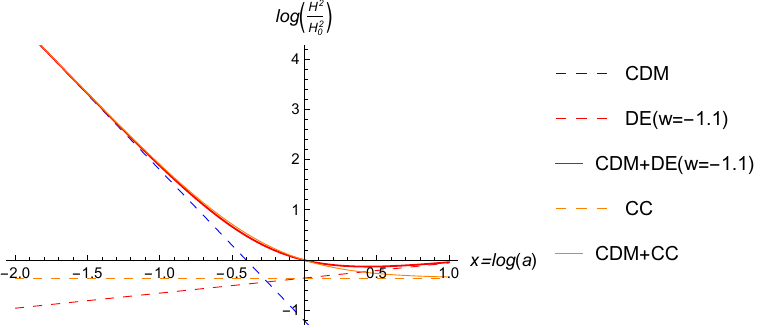}
  \caption{``CDM + DE ($w=-1.1$)" is an example with $\Omega_m=0.3$, $\Omega_{\rm DE}=1-\Omega_m$ and EOS of DE $w<-1$. ``CDM + CC" is the standard model with $\Omega_m=0.3$, $\Omega_{\rm DE}=1-\Omega_m$.
  The difference between ``CDM + DE ($w=-1.1$)" and ``CDM + CC" is small for $x\le 0$ in this figure and is amplified in Fig.~\ref{fig:DE_wSmaller_DIFF}.
  }
  \label{fig:DE_wSmaller}
  \vspace{1cm}
  \includegraphics[width=0.7\textwidth]{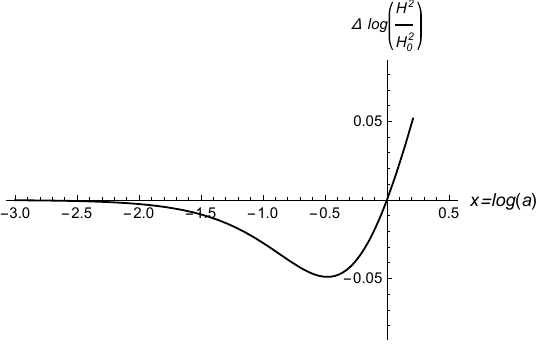}
  \caption{The difference between ``CDM + DE ($w=-1.1$)" and ``CDM + CC" in Fig.~\ref{fig:DE_wSmaller}.}
  \label{fig:DE_wSmaller_DIFF}
\end{figure}

\begin{figure}[H]
  \centering
  \includegraphics[width=0.9\textwidth]{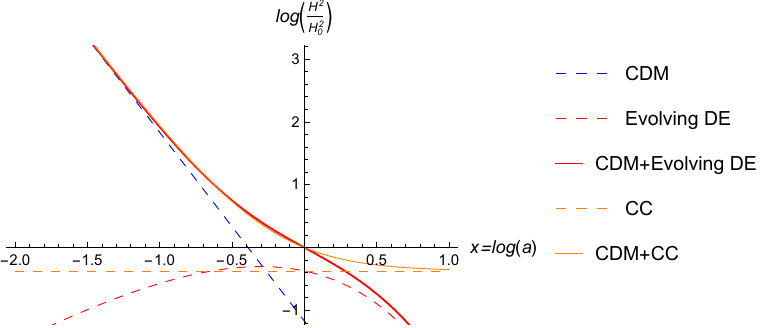}
  \caption{``CDM + Evolving DE" is the DESI+CMB+Pantheon+ best-fit model with $\Omega_m=0.3114$, $\Omega_{\rm DE}=1-\Omega_m$, $w_0=-0.838$, $w_a=-0.62$. ``CDM + CC" is the standard model with $\Omega_m=0.3114$, $\Omega_{\rm DE}=1-\Omega_m$.
  The difference between ``CDM + Evolving DE" and ``CDM + CC" is small for $x\le 0$ in this figure and is amplified in Fig.~\ref{fig:DESI_DE_DIFF}
  }
  \label{fig:DESI_DE}
  \vspace{1cm}
  \centering
  \includegraphics[width=0.7\textwidth]{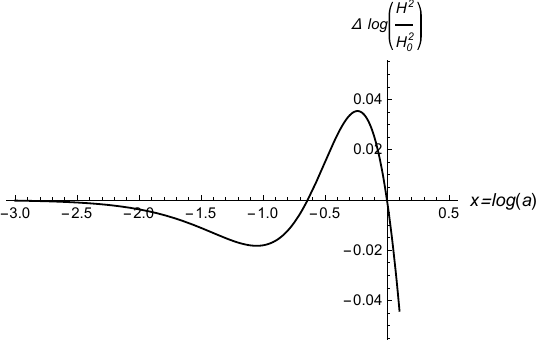}
  \caption{The difference between ``CDM + Evolving DE" and ``CDM + CC" in Fig.~\ref{fig:DESI_DE}. This is the rough shape of the Hubble parameter evolution that leads to the claim of evolving dark energy.}
  \label{fig:DESI_DE_DIFF}
\end{figure}

In Fig.~\ref{fig:DE_wLarger}--\ref{fig:DESI_DE_DIFF}, we show a few examples of the evolution of $H$ in the non-standard, including evolving, dark energy scenarios. For illustration purpose, except for the cases we explicitly mention, such as the DESI best fits, the parameters in examples do not necessarily take the same values as those of the standard model. We pay attention to the evolution of $H$ up to today $a=a_0=1$, although in all figures we also show a portion of extension beyond today (these extensions are modifiable and unconstrained).

As we can see, in all cases, dark energy redshifts faster than CDM, so its effect at the very early universe is negligible. If $w_{\rm DE}$ is larger than $-1$, the dark energy redshifts faster than a CC, leading to a late-time excess when compared with the standard model (Fig.~\ref{fig:DE_wLarger} and \ref{fig:DE_wLarger_DIFF}), and we see a bump in Fig.~\ref{fig:DE_wLarger_DIFF}. Such a qualitative behavior also holds for quintessence-like models in which $w_{\rm DE}$ is evolving but remains larger than -1. For cases where $w_{\rm DE}<-1$, which is the NEC violating case, the situation is the opposite (Fig.~\ref{fig:DE_wSmaller} and \ref{fig:DE_wSmaller_DIFF}), and we see a dip in Fig.~\ref{fig:DE_wSmaller_DIFF}.

{\em The DESI DR2 case.} Interestingly, the best-fit model from DESI+CMB+Pantheon+ observations currently suggests a dark energy model of which the EOS is transiting between the two behaviors (Fig.~\ref{fig:DESI_DE} and \ref{fig:DESI_DE_DIFF}), and we see an oscillation in Fig.~\ref{fig:DESI_DE_DIFF}. Fig.~\ref{fig:DESI_DE_DIFF} shows the rough shape of the Hubble parameter evolution that underlies the anomaly observed by DESI and Pantheon+ and leads to the type of evolving dark energy in the main claim of DESI DR2.

\section{Evolving dark matter}
\label{Sec:Evolving DM}

Now we explore the possibility in which the dark energy is still a cosmological constant, but the deviation from the standard model observed in Fig.~\ref{fig:DESI_DE_DIFF} is due to assumptions in the matter sector. We would like to introduce an exotic dark matter (EDM) component whose EOS is non-standard but does not violate the NEC. In order to satisfy the early universe constraints, the EOS of the EDM has to be very close to zero, so the non-standard EOS should be introduced at late time, namely, in a time-dependent fashion. We will soon see that, in order to fit the property such as that in Fig.~\ref{fig:DESI_DE_DIFF}, it also has to be oscillatory. On the way to the examples most relevant to DESI DR2, we also demonstrate a variety of other examples which may help understand the increasingly complicated examples and which could be useful in different contexts.

To make physics more transparent and expressions simpler, we model the time-dependence in the EOS of the exotic dark matter, $w_{\rm EDM}$, using piece-wise step functions. The main purpose is to explore different possibilities. We leave more elaborated modeling to future works.

\subsection{One step function case}

Let us first investigate the case in which $w_{\rm EDM}$ deviates from 0 to either negative values or positive values, starting from $a=a_c$. We study the qualitative features of such models by introducing one step function,
\begin{align}
    w_{\rm EDM} (a) = 
\begin{cases}
0 & \text{if } a < a_{c} ~,\\
w_1 & \text{if } a\ge a_{c} ~.
\end{cases}
\end{align}
The evolution of the Hubble parameter is then given by
\begin{align}
    \frac{H^2}{H_0^2} = 
    \begin{cases}
    \Omega_m a^{-3} + \Omega_{\rm EDM}~ a_c^{-3w_1} a^{-3} + \Omega_{\rm DE} 
    & \text{if } a < a_{c} ~,\\
    \Omega_m a^{-3} + \Omega_{\rm EDM}~ a^{-3(1+w_1)} + \Omega_{\rm DE} 
    & \text{if } a\ge a_{c} ~.
    \end{cases}
    \label{eq:Friedmann_EDM}
\end{align}
For a flat universe, $\Omega_m+\Omega_{\rm EDM}+\Omega_{\rm DE}=1$.
Due to the late-time modification from the exotic dark matter, a LCDM model, which should be used to match the early time, matter-dominated, evolution of Eq.~\eqref{eq:Friedmann_EDM}, does not necessarily has its matter density parameter given by $\Omega_m+\Omega_{\rm EDM}$. In general, a constant shift is needed.

\begin{figure}[H]
  \centering
  \includegraphics[width=0.9\textwidth]{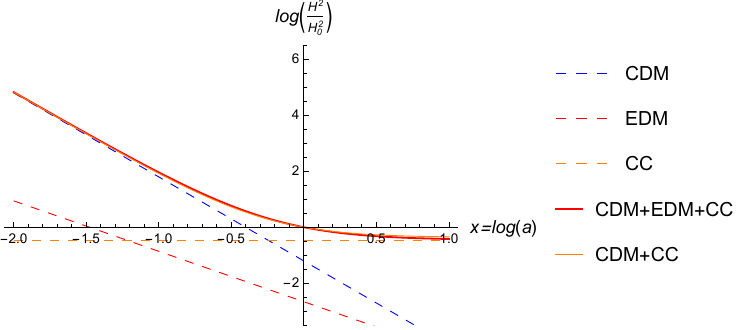}
  \caption{
  For the model with exotic DM, ``CDM+EDM+CC", the parameters are: $\Omega_m=0.3$, $\Omega_{\rm EDM}=0.07$, $\Omega_{\rm CC}=1-0.3-0.07$.
  For the LCDM model, ``CDM + CC", a shift in the density parameters is added and adjusted to diminish the early-time difference between the two models: $\Omega_m=0.3+0.07-0.064$, $\Omega_{\rm CC}=1-0.3-0.07+0.064$.
  The evolution of $w_{\rm EDM}$ is plotted in Fig.~\ref{fig:EDM_single_bump_wEDM}.
  The difference between ``CDM+EDM+CC" and ``CDM + CC" is too small to see clearly in this figure, and we amplify it in Fig.~\ref{fig:EDM_single_bump_diff}.
  }
  \label{fig:EDM_single_bump}
  \vspace{1cm}
  \includegraphics[width=0.7\textwidth]{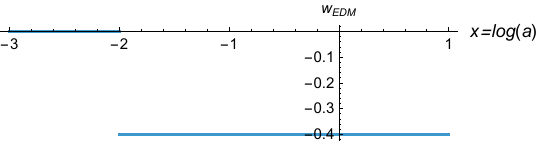}
  \caption{The evolution of $w_{\rm EDM}$.}
  \label{fig:EDM_single_bump_wEDM}
  \vspace{1cm}
  \includegraphics[width=0.7\textwidth]{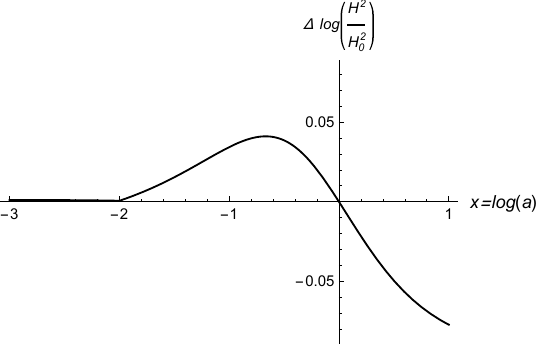}
  \caption{The difference between ``CDM+EDM+CC" and ``CDM + CC" in Fig.~\ref{fig:EDM_single_bump}.
  Such an EDM model could be confused with a non-standard dark energy model in Fig.~\ref{fig:DE_wLarger_DIFF}.
  }
  \label{fig:EDM_single_bump_diff}
\end{figure}

\begin{figure}[H]
  \centering
  \includegraphics[width=0.9\textwidth]{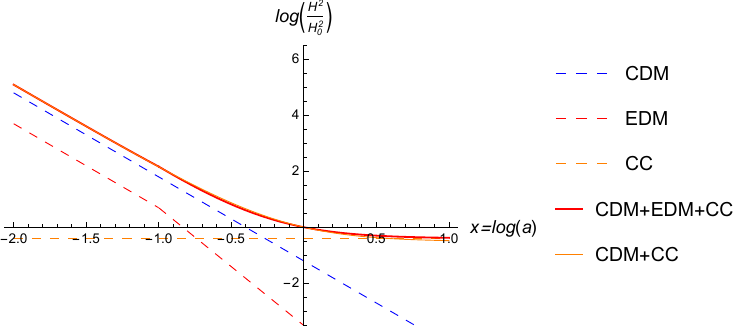}
  \caption{
  For the model with exotic DM, ``CDM+EDM+CC", the parameters are: $\Omega_m=0.3$, $\Omega_{\rm EDM}=0.03$, $\Omega_{\rm CC}=1-0.3-0.03$.
  For the LCDM model, ``CDM + CC", a shift in the density parameters is added and adjusted to diminish the early-time difference between the two models: $\Omega_m=0.3+0.03+0.07$, $\Omega_{\rm CC}=1-0.3-0.03-0.07$.
  }
  \label{fig:EDM_single_dip}
  \vspace{1cm}
  \includegraphics[width=0.7\textwidth]{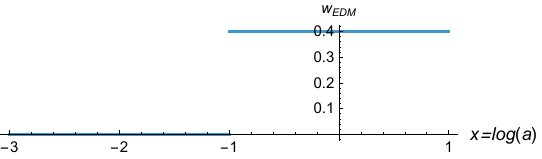}
  \caption{The evolution of $w_{\rm EDM}$}
  \label{fig:EDM_single_dip_wEDM}
  \vspace{1cm}
  \includegraphics[width=0.7\textwidth]{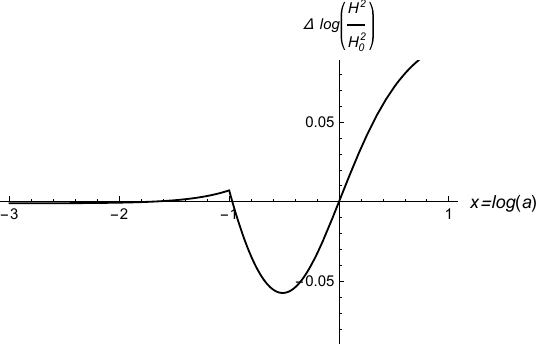}
  \caption{The difference between ``CDM+EDM+CC" and ``CDM + CC" in Fig.~\ref{fig:EDM_single_dip}. Such an EDM model could be confused with a phantom dark energy model in Fig.~\ref{fig:DE_wSmaller_DIFF}.}
  \label{fig:EDM_single_dip_diff}
\end{figure}

In Fig.~\ref{fig:EDM_single_bump}, \ref{fig:EDM_single_bump_wEDM}, \ref{fig:EDM_single_bump_diff} and Fig.~\ref{fig:EDM_single_dip}, \ref{fig:EDM_single_dip_wEDM}, \ref{fig:EDM_single_dip_diff}, we demonstrate two such examples.

The differences between the new models and the LCDM models demonstrated in these examples can be readily understood. If some negative values of $w_{\rm EDM}$ is turned on at a late time, this exotic dark matter component starts to be red-shifted slower than cold dark matter, so the Hubble parameter decreases more slowly. Because we fix today's Hubble parameter to be $H_0$ and match the early time behavior to a LCDM model, we see a bump feature in $\log(H_{\rm new \, model}^2)-\log(H_{\rm LCDM}^2)$. In the case where some positive values of $w_{\rm EDM}$ is turned on, we see a dip feature. So, we already start to see that, within certain experimental sensitivities, observationally, exotic dark matter could be confused with non-standard dark energy illustrated in examples in Fig.~\ref{fig:DE_wLarger}, \ref{fig:DE_wLarger_DIFF} and Fig.~\ref{fig:DE_wSmaller}, \ref{fig:DE_wSmaller_DIFF}. 
As mentioned, although, in these latter figures (the CDM+DE models), the $w_{\rm DE}$'s of the dark energy are constants, the demonstrated deviations from the LCDM model are qualitatively the same for other dark energy models where the $w_{\rm DE}$'s slowly vary with time but remain above $-1$ (such as the quintescence model), or remain below $-1$, respectively.

On the other hand, such Hubble parameter evolutions are not the one specifically highlighted by DESI DR2 (but some could still be statistically allowed by DESI DR2), so let us move on to more complicated possibilities in which $w_{\rm EDM}$ oscillates between positive and negative values after being turned on at a late time.

\subsection{Two step-function case}

\begin{figure}[H]
  \centering
  \includegraphics[width=0.9\textwidth]{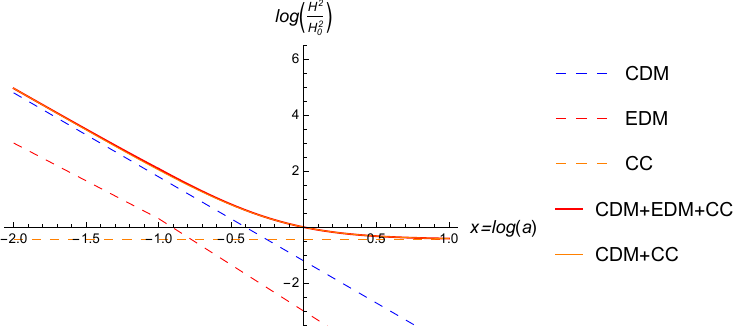}
  \caption{
  For the model with exotic DM, ``CDM+EDM+CC", the parameters are: $\Omega_m=0.3$, $\Omega_{\rm EDM}=0.05$, $\Omega_{\rm CC}=1-0.3-0.05$.
  For the LCDM model, ``CDM + CC", the parameters are: $\Omega_m=0.3+0.05$, $\Omega_{\rm CC}=1-0.3-0.05$.
  }
  \label{fig:EDM_double_bump}
  \vspace{1cm}
  \includegraphics[width=0.7\textwidth]{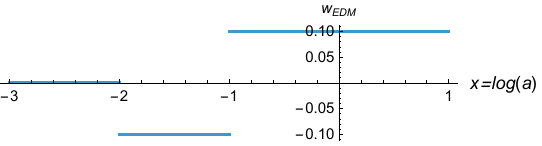}
  \caption{The evolution of $w_{\rm EDM}$.}
  \label{fig:EDM_double_bump_wEDM}
  \vspace{1cm}
  \includegraphics[width=0.7\textwidth]{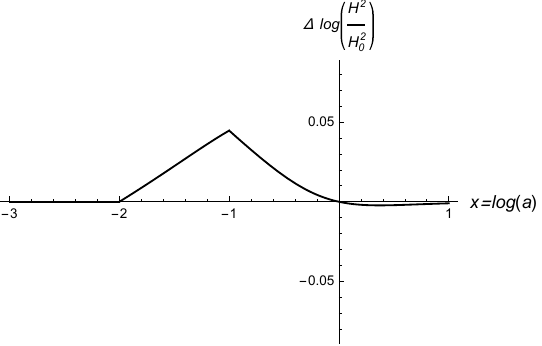}
  \caption{The difference between ``CDM+EDM+CC" and ``CDM + CC" in Fig.~\ref{fig:EDM_double_bump}.}
  \label{fig:EDM_double_bump_diff}
\end{figure}

\begin{figure}[H]
  \centering
  \includegraphics[width=0.9\textwidth]{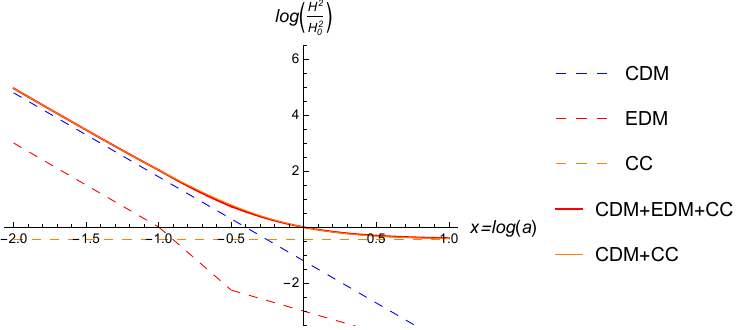}
  \caption{
  For the model with exotic DM, ``CDM+EDM+CC", the parameters are: $\Omega_m=0.3$, $\Omega_{\rm EDM}=0.05$, $\Omega_{\rm CC}=1-0.3-0.05$.
  For the LCDM model, ``CDM + CC", the parameters are: $\Omega_m=0.3+0.05$, $\Omega_{\rm CC}=1-0.3-0.05$.
  }
  \label{fig:EDM_double_dip}
  \vspace{1cm}
  \includegraphics[width=0.7\textwidth]{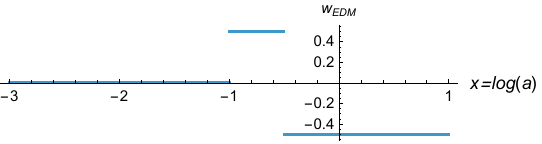}
  \caption{The evolution of $w_{\rm EDM}$.}
  \label{fig:EDM_double_dip_wEDM}
  \vspace{1cm}
  \includegraphics[width=0.7\textwidth]{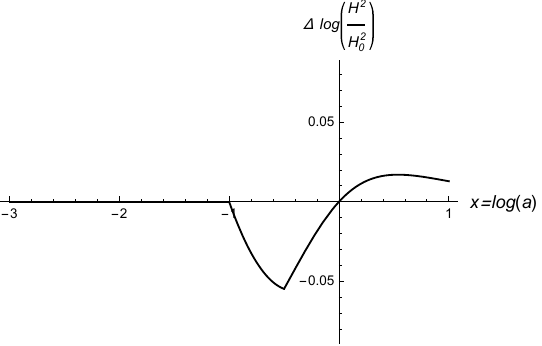}
  \caption{The difference between ``CDM+EDM+CC" and ``CDM + CC" in Fig.~\ref{fig:EDM_double_dip}.}
  \label{fig:EDM_double_dip_diff}
\end{figure}

\begin{figure}[H]
  \centering
  \includegraphics[width=0.9\textwidth]{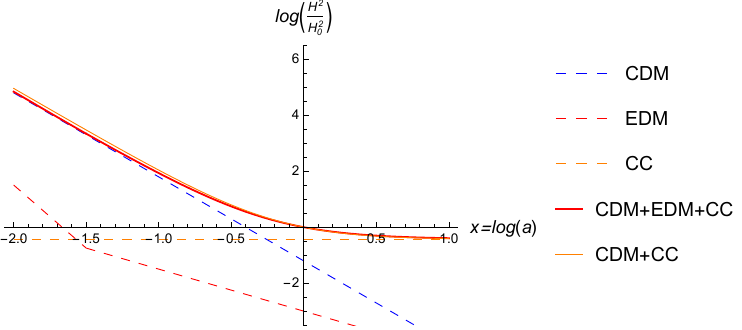}
  \caption{
  For the model with exotic DM, ``CDM+EDM+CC", the parameters are: $\Omega_m=0.3$, $\Omega_{\rm EDM}=0.05$, $\Omega_{\rm CC}=1-0.3-0.05$.
  For the LCDM model, ``CDM + CC", a shift in the density parameters is added and adjusted to diminish the early-time difference between the two models: $\Omega_m=0.3+0.05-0.039$, $\Omega_{\rm CC}=1-0.3-0.05+0.039$.
  }
  \label{fig:EDM_double_osci}
  \vspace{1cm}
  \includegraphics[width=0.7\textwidth]{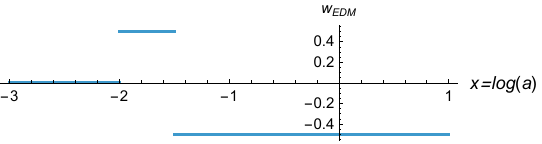}
  \caption{The evolution of $w_{\rm EDM}$.}
  \label{fig:EDM_double_osci_wEDM}
  \vspace{1cm}
  \includegraphics[width=0.7\textwidth]{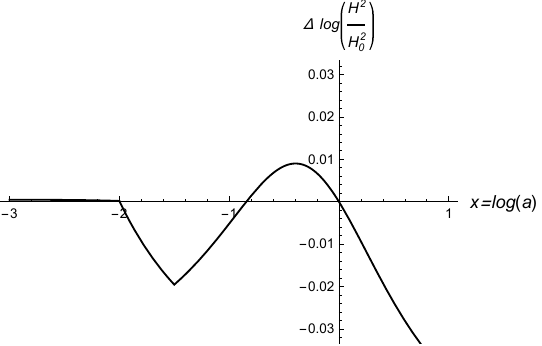}
  \caption{The difference between ``CDM+EDM+CC" and ``CDM + CC" in Fig.~\ref{fig:EDM_double_osci}.}
  \label{fig:EDM_double_osci_diff}
\end{figure}

\begin{figure}[H]
  \centering
  \includegraphics[width=0.9\textwidth]{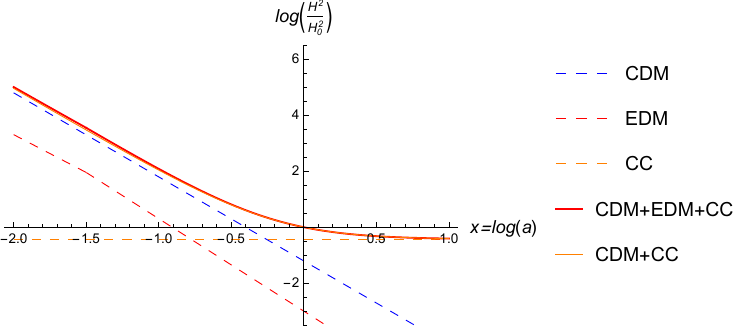}
  \caption{
  For the model with exotic DM, ``CDM+EDM+CC", the parameters are: $\Omega_m=0.3$, $\Omega_{\rm EDM}=0.05$, $\Omega_{\rm CC}=1-0.3-0.05$.
  For the LCDM model, ``CDM + CC", a shift in the density parameters is added and adjusted to diminish the early-time difference between the two models: $\Omega_m=0.3+0.05+0.0175$, $\Omega_{\rm CC}=1-0.3-0.05-0.0175$.
  }
  \label{fig:EDM_double_osci_2}
  \vspace{1cm}
  \includegraphics[width=0.7\textwidth]{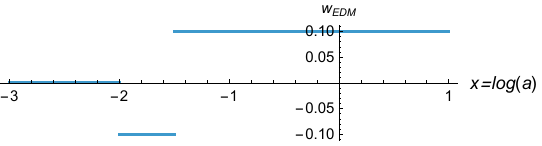}
  \caption{The evolution of $w_{\rm EDM}$.}
  \label{fig:EDM_double_osci_wEDM_2}
  \vspace{1cm}
  \includegraphics[width=0.7\textwidth]{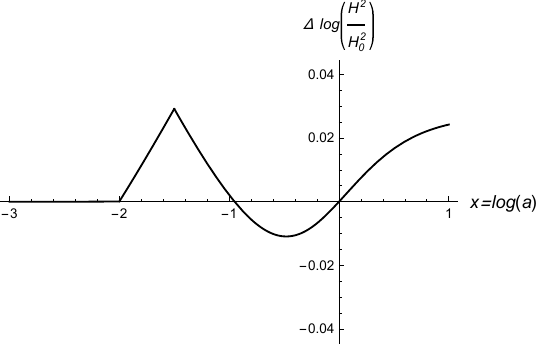}
  \caption{The difference between ``CDM+EDM+CC" and ``CDM + CC" in Fig.~\ref{fig:EDM_double_osci_2}.}
  \label{fig:EDM_double_osci_diff_2}
\end{figure}

Next, we consider the case of oscillating $w_{\rm EDM}$. We model the oscillation in $w_{\rm EDM}$ by two step functions consecutively,
\begin{align}
    w_{\rm EDM} (a) = 
\begin{cases}
0 & \text{if } a < a_{c1} ~,\\
w_1 & \text{if } a_{c1} \le a< a_{c2} ~, \\
w_2 & \text{if } a \ge a_{c2} ~.
\end{cases}
\end{align}
The results are show in Fig.~\ref{fig:EDM_double_bump}, \ref{fig:EDM_double_bump_wEDM}, \ref{fig:EDM_double_bump_diff}; Fig.~\ref{fig:EDM_double_dip}, \ref{fig:EDM_double_dip_wEDM}, \ref{fig:EDM_double_dip_diff}; Fig.~\ref{fig:EDM_double_osci}, \ref{fig:EDM_double_osci_wEDM}, \ref{fig:EDM_double_osci_diff}; and Fig.~\ref{fig:EDM_double_osci_2}, \ref{fig:EDM_double_osci_wEDM_2}, \ref{fig:EDM_double_osci_diff_2}.

We can see that having an oscillating $w_{\rm EDM}$ leads to more flexibility. 

Because $w_{\rm EDM}$ oscillates between a positive and a negative value, it is possible that their net effect brings the changes in $H(a)$ back to a LCDM model with the matter density parameter equal to $\Omega_m+\Omega_{\rm EDM}$; so an extra shift may not be needed in this parameter of the LCDM model used as the baseline model. The two examples, Fig.~\ref{fig:EDM_double_bump}, \ref{fig:EDM_double_bump_wEDM}, \ref{fig:EDM_double_bump_diff}, and Fig.~\ref{fig:EDM_double_dip}, \ref{fig:EDM_double_dip_wEDM}, \ref{fig:EDM_double_dip_diff}, are such examples. In these examples, bump or dip features similar to some previous cases are created, but due to the presence of two step functions, the height and shapes of these bump or dip features are more tunable.

On the other hand, in general the effects of positive and negative valued $w_{\rm EDM}$ on $H(a)$ do not cancel. In these cases, when choosing a baseline LCDM model, a shift (from $\Omega_m+\Omega_{\rm EDM}$ given by the new model) is needed when adjusting the matter density parameter of this LCDM model. Once this is done, we start to see various interesting oscillating features in $\log(H_{\rm new \, model}^2)-\log(H_{\rm LCDM}^2)$, for example, in Fig.~\ref{fig:EDM_double_osci}, \ref{fig:EDM_double_osci_wEDM}, \ref{fig:EDM_double_osci_diff}; and Fig.~\ref{fig:EDM_double_osci_2}, \ref{fig:EDM_double_osci_wEDM_2}, \ref{fig:EDM_double_osci_diff_2}. 
In particular, the case Fig.~\ref{fig:EDM_double_osci}, \ref{fig:EDM_double_osci_wEDM}, \ref{fig:EDM_double_osci_diff} generates an oscillating feature close to that suggested by DESI DR2 (see Fig.~\ref{fig:DESI_DE}, \ref{fig:DESI_DE_DIFF}).

The value of the shift in matter density parameter mentioned above is constrained by other values of the models. The advantage of this constraint is that fewer parameter is needed; the disadvantage is that the shape and amplitude of the oscillating feature could be less flexible. So, to explore more possibilities, let us move on to the examples where $w_{\rm EDM}$ has more oscillation.

\subsection{Three step-function case}

Let us now add one more oscillation in $w_{\rm EDM}$,
\begin{align}
    w_{\rm EDM} (a) = 
\begin{cases}
0 & \text{if } a < a_{c1} ~,\\
w_1 & \text{if } a_{c1} \leq a < a_{c2} ~,\\
w_2 & \text{if } a_{c2} \leq a < a_{c3} ~,\\
w_3 & \text{if } a \geq a_{c3} ~.
\end{cases}
\end{align}

\begin{figure}[H]
  \centering
  \includegraphics[width=0.9\textwidth]{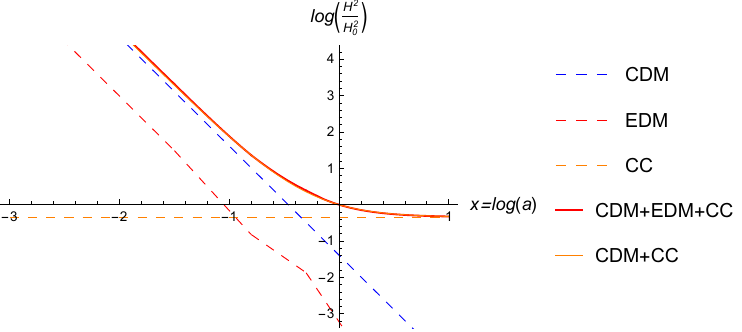}
  \caption{
  For the model with exotic DM, ``CDM+EDM+CC", the parameters are: $\Omega_m=0.25$, $\Omega_{\rm EDM}=0.04$, $\Omega_{\rm CC}=1-0.25-0.04$.
  For the LCDM model, ``CDM + CC", a shift in the density parameters is added and adjusted to diminish the early-time difference between the two models: $\Omega_m=0.25+0.04+0.0095$, $\Omega_{\rm CC}=1-0.25-0.04-0.0095$.
  The evolution of $w_{\rm EDM}$ is plotted in Fig.~\ref{fig:EDM_triple_osci_wEDM}.
  The difference between ``CDM+EDM+CC" and ``CDM + CC" is too small to see clearly in this figure, and we amplify it in Fig.~\ref{fig:EDM_triple_osci_diff}.
  }
  \label{fig:EDM_triple_osci}
  \vspace{1cm}
  \includegraphics[width=0.7\textwidth]{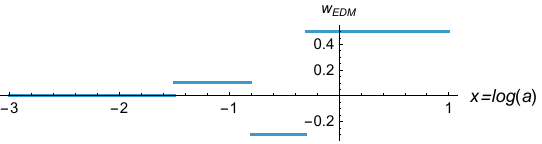}
  \caption{The evolution of $w_{\rm EDM}$.}
  \label{fig:EDM_triple_osci_wEDM}
  \vspace{1cm}
  \includegraphics[width=0.7\textwidth]{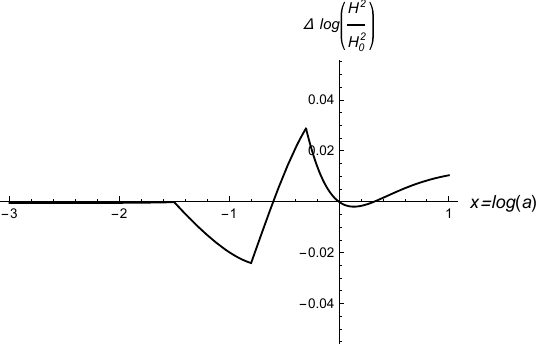}
  \caption{The difference between ``CDM+EDM+CC" and ``CDM + CC" in Fig.~\ref{fig:EDM_triple_osci}. Such a model is capable of reproducing the features in Fig.~\ref{fig:DESI_DE}, \ref{fig:DESI_DE_DIFF}.}
  \label{fig:EDM_triple_osci_diff}
\end{figure}

With more oscillation and parameters, the feature can now be generated with more flexibilities. We will not try to list all possibilities here, but simply show an example in Fig.~\ref{fig:EDM_triple_osci}, \ref{fig:EDM_triple_osci_wEDM}, \ref{fig:EDM_triple_osci_diff} to demonstrate that such a case is capable of generating an oscillating feature very close to that of the evolving dark energy model highlighted by DESI DR2, for $a<1$ and within the current experimental precision. Such a feature, entirely generated by the dark matter sector, which is non-gravitationally decoupled from the dark energy sector, contains an exotic dark matter species with an oscillating EOS, and exhibits no obvious pathological property, could be misinterpreted as an evolving dark energy component that violates the NEC condition, if we model the entire dark matter sector as the standard cold dark matter in data-fitting.

\section{An example of evolving dark matter}
\label{Sec:EDM example}

Here we give an explicit example for the underlying physics of exotic dark matter whose EOS  changes with time. 
We model this species of dark matter by a scalar field $\phi$ with the Lagrangian density 
\begin{align}
    {\cal L} = \frac{1}{2} \partial_\mu\phi \partial^\mu\phi - V(\phi) ~,
    \label{Eq:L_phi}
\end{align}
and treat it as a coherent state of this field theory.
The field $\phi$ undergoes rapid and coherent oscillations around the minimum of the potential $V(\phi)$, and when averaged over timescales much longer than the oscillation period, it behaves effectively as a fluid. The oscillation period, $T$, is orders of magnitude shorter than the time scale of cosmic consideration, $T\ll 1/H$, as well as time scales of most astrophysical phenomena.

The EOS of such a fluid is determined by the shape of the potential the field $\phi$ is probing. For example, for a harmonic potential $\sim\phi^2$, it has $w=0$; for a constant potential, it is a cosmological constant ($w=-1$); and, more generally, for $V=\lambda \phi^n$, because at the background level $\langle \dot\phi^2 \rangle= n\lambda \langle \phi^n \rangle$, we have
\begin{align}
    w=\frac{n-2}{n+2} ~.
\end{align}

As the universe expands, the oscillation amplitude of $\phi$ gets Hubble-damped. If the shape of the potential is complicated enough, with time, the overall shape of the lower part of the potential, that $\phi$ is probing at later times, may be different from that of the upper part, that it is probing at earlier times. Effectively, the EOS of this fluid is changing with time.

To give an explicit example of such a potential, let us reverse-engineer a potential that gives rise to the evolution of $w_{\rm EDM}$ in Fig.~\ref{fig:EDM_double_osci_wEDM}:
\begin{align}
    V(\phi) \propto \frac{f(\phi)^6}{\phi_1^4 + f(\phi)^4} ~, ~~~ {\rm where} ~~
    f(\phi)=\left( \phi_2^{16/3} + |\phi|^{16/3} \right)^{1/6} |\phi|^{1/9} ~,
    \label{Eq:V_example}
\end{align}
and $\phi_1$ and $\phi_2$ are two constant parameters satisfying $\phi_1 \gg \phi_2>0$.
As one can check, the overall shapes of the potential behave as
\begin{align}
    V(\phi) \propto
    \begin{cases}
        \phi^2 ~, & |\phi|\gg \phi_1 ~, \\
        \phi^6 ~, & \phi_1 \gg |\phi| \gg \phi_2 ~, \\
        \phi^{2/3} ~, & \phi_2 \gg |\phi| ~.
    \end{cases}
\end{align}
So, as the universe expands, the background fluid first behaves as a CDM, then with $w_{\rm EDM} \approx 0.5$, then  $w_{\rm EDM} \approx -0.5$. The values of $\phi_{1,2}$ will be determined by comparing the detail of the $\phi$ field evolution with observational data on the Hubble parameter evolution.
Interestingly, in this scenario, DESI is effectively probing the shape of the potential landscape of the dark matter field, instead of properties of dark energy.

As a candidate for exotic dark matter species, it is also important to examine its properties at the perturbation level under gravitation, especially its sound speed, $c_s$, and associated Jeans length, $\lambda_J$. 

We first note that $c_s^2$ is defined at the perturbation level and is generally not the same as $w$ (unless $P$ is a function of only the background density $\rho$). So, $c_s^2$ can be zero or positive, even if $w$ is negative. 

We also note that, although it is well-known that for a scalar field theory with a canonical kinetic term such as \eqref{Eq:L_phi}, the sound speed is 1 \cite{Garriga:1999vw}, this sound speed is not necessarily the same quantity of interest in the analyses of gravitational instability. In the latter context, the sound speed is 1 only in the relativistic regime, where the physical wavenumber, $k/a$, is much larger than the energy scale of the oscillation, $1/T$, $k/a\gg 1/T$. In the non-relativistic regime, the gravitational instability can act before the effective scalar particles escape the overdense region. This scale can be estimated as follows. For a mode with a physical wavelength $\lambda$, we can estimate the time for the particle to escape the overdense region as
\begin{align}
    t_{\rm esc} \sim \frac{\lambda}{v} \sim \frac{2\pi a/k}{(k/a)/(2\pi/T)} \sim \frac{a^2}{k^2 T} ~,
\end{align}
in which we have associated an effective mass $\sim 2\pi/T$ to an effective particle for a scalar field oscillating with the period $T$. On the other hand, the dynamical time for gravity is given by 
\begin{align}
    t_{\rm dyn} \sim (4\pi G\rho)^{-1/2} \sim H^{-1} ~.
\end{align}
The particles collapse under gravity if $t_{\rm dyn} < t_{\rm esc}$, namely, on scales $a/k>\sqrt{T/H}$. This corresponds to a Jeans length $\lambda_J\sim \sqrt{T/H}$. Rigorous derivation of this scale with a harmonic potential ($T=2\pi/m$) has been derived in the context of the axion dark matter (e.g.~\cite{Sikivie:2009qn}).
Because $T$ is orders of magnitude smaller than $1/H$, this length scale is much shorter than the Hubble length. So, in a vast range of scales under the Hubble horizon, this scalar field can have effectively zero sound speed (as well as model-dependently tunable) despite the fact that its equation of state could undergo significant time evolution. In particular, in the early universe, by construction, we have $w=0$, this dark matter species effectively behaves as the standard cold dark matter, making it unconstrained by the CMB physics; however, for cases where the energy budget among components need to be slightly adjusted with respect to the standard model, a careful reanalysis in early universe physics will be needed.

Another interesting property is that the EOS of this EDM can also be mode- and density-dependent. In the main part of this paper, the zeroth mode of the EDM is used for the background evolution. Shorter wavelength modes are relevant for structure formation, and in higher density regions, these modes have larger oscillation amplitudes in their coherent states. As a result, these modes probe different regions of the potential landscape compared to the zeroth mode. For example, in \eqref{Eq:V_example}, the higher the local density, the closer is the EOS of EDM to zero.

This is only one example of evolving DM. It would be interesting to explore other examples.

\section{Other constraints on evolving dark matter}
\label{Sec:Constraints}

The examples in Fig.~\ref{fig:EDM_double_osci}, \ref{fig:EDM_double_osci_wEDM}, \ref{fig:EDM_double_osci_diff} and Fig.~\ref{fig:EDM_triple_osci}, \ref{fig:EDM_triple_osci_wEDM}, \ref{fig:EDM_triple_osci_diff} seems to suggest an exotic dark matter species that has a density parameter of order $10-20\%$ of the rest of the matter content (e.g. $\Omega_{\rm EDM}\sim 0.04$) and a EOS oscillating between $\pm 0.1\sim \pm 0.5$ may be needed. On the other hand, the detailed sequence of how $\Omega_{\rm EDM}$ evolves can be very different, even though the consequences on $H(a)$ (alone) are close. For example, in the example of Fig.~\ref{fig:EDM_double_osci}, \ref{fig:EDM_double_osci_wEDM}, \ref{fig:EDM_double_osci_diff}, $w_{\rm EDM}$, as a function of redshift $z$, evolves roughly in the following way:
\begin{align}
    w_{\rm EDM} \sim
\begin{cases}
0~, & z > 6.4 ~,\\
0.5~, & 6.4 \leq z < 3.5 ~,\\
-0.5~, & 3.5 \leq z < 0 ~;
\end{cases}
\end{align}
while in the example of Fig.~\ref{fig:EDM_triple_osci}, \ref{fig:EDM_triple_osci_wEDM}, \ref{fig:EDM_triple_osci_diff},
\begin{align}
    w_{\rm EDM} \sim
\begin{cases}
0~, & z > 3.5 ~,\\
0.1 ~, & 3.5 \leq z < 1.2 ~,\\
-0.3 ~, & 1.2 \leq z < 0.35 ~,\\
0.5~, & 0.35 \leq z < 0 ~.
\end{cases}
\end{align}

So far, we have only considered the effect of such EDM on the evolution of the Hubble parameter. EDM with a non-standard and/or evolving EOS and a non-zero sound speed could also impact a variety of astrophysical phenomena (see e.g.~\citep{Hu:1998kj,Kopp:2018zxp,Pop_awski_2019,Davari:2023tam}), such as galaxy formation, their halo density profiles, or growth of structure. 

Although the pressure of EDM does not play an important role in the initial linear perturbative evolution, it can become important in astrophysical hydrodynamics later on. If its EOS is very different from zero, it could lead to important constraints. On the other hand, this is a very model-dependent subject. For example, in the model of Sec.~\ref{Sec:EDM example}, as discussed, the relevant values of EOS in astrophysics and in background cosmology are very different. In background cosmology, $w_{\rm EDM}$ can deviate from 0 as late as near $z=5$, at which the average density of the universe is around $\sim 10^{-25} {\rm kg}~{\rm m}^{-3}$, much smaller than values that are of interests in most astrophysical phenomena. So, at least in this model, EDM locally behaves very close to CDM even though its cosmological average is not.

Properties of such an EDM relevant to the background cosmology may be studied in regions between galaxies and clusters where the dark matter density is very diffused -- close to the average density of the universe today. Interestingly, if we could study their properties in cosmic voids where the density is lower than the average, we could even predict the future impact of this EDM on the background cosmology.

The effects of EDM's sound speed, the derivative of the pressure with respect to density, can also be very model-dependent. 
If the sound speed is practically zero, the EDM will stay confined in galaxies and clusters of galaxies. However, if it is finite, the EDM cannot fall into shallow gravitational potential wells with an escape speed below the sound speed. The escape speed ranges from $\sim 10^{-4}c=30~{\rm km~s^{-1}}$ for dwarf galaxies and up to $\sim 10^{-2}c=3,000~{\rm km~s^{-1}}$ for massive clusters of galaxies. But even when EDM remains unbound, if it only consists of $\sim 15\%$ of the dark matter mass budget, its influence on the mass profiles of gravitationally-bound systems would be minor and difficult to ascertain with current data. If the sound speeds take intermediate values, it may be interesting to explore whether such EDM could modify the halo density profile, particularly in the cusp region.

Other types of astrophysical observations that could potentially be used to test or constrain such EDM include studies of their effects on the growth rate of structure, as well as CMB and galaxy lensing.

Although DESI DR2 constraint on $H_0$, assuming evolving dark energy, is very close to that measured by CMB, a generic consequence of having EDM (as well as evolving dark energy) is to modify this value, as they modify the late-time evolution of cosmology. As we have demonstrated, there are a variety of possibilities in the evolution of EDM; it would be interesting to see if any of these could be used to fit the current, or future, data, and at the same time relieve the Hubble tension \cite{DiValentino:2021izs}.

\section{Conclusions}
\label{Sec:Conclusions}

Our quantitative modeling and analysis show that the latest empirical constraints on cosmology from a combination of DESI, CMB and supernova data can be accounted for if $\sim 15\%$ of dark matter has an evolving and oscillating equation of state within $-1<w<1$. From a fundamental physics perspective, this interpretation is more appealing than an evolving phantom dark energy with $w<-1$, which violates the null energy condition. 

It would be interesting to construct smoother templates, as well as more explicit models, of this kind and compare them with different data sets. Some quantitative features of these models may shift when fitted to specific data sets or updated in light of future observations. Since the predictions from the evolving dark matter and evolving dark energy cannot be identical, even if they are currently both allowed empirically, future data or other constraints may be able to tell the difference.

In addition to examples that are most relevant to the recent DESI DR2 result, we have also demonstrated several other examples of effects evolving dark matter can have on cosmology.
The ideas and methods presented here could also be useful if future data points in a different direction.

\bigskip

{\bf Notes added:} During the preparation of this paper, a couple papers appeared which constrain non-standard dark matter with non-zero equation of state using the CMB and DESI data. Ref.~\cite{Kumar:2025etf} uses a constant non-zero $w$ and Ref.~\cite{Wang:2025zri} uses a linear ansatz, $w_{\rm dm}= w_{\rm dm0} + w_{\rm dma}(1-a)$. (Also see earlier works on this aspect \cite{Kopp:2018zxp}.) The difference of these studies from our work is that these ansatz do not necessarily aim to reinterpret the evolving, especially phantom, nature of the dark energy claimed in DESI DR2, which is the main focus of our Sec.~\ref{Sec:Evolving DM}. We demonstrated that an oscillating $w$ is needed for this purpose. A constant or linear ansatz can at most capture some averaged effects of such an oscillation. In retrospect, it is also readily understood that the evidence found for the non-standard dark matter of these works is much smaller than that found in DESI DR2 for evolving dark energy.

\medskip
\section*{Acknowledgments}

We would like to thank Daniel Eisenstein and Cristhian Garcia Quintero for discussions on the DESI results and this paper. AL was supported in part by the Black Hole Initiative, which is funded by GBMF anf JTF.



\bibliographystyle{JHEP}
\bibliography{Ref}

\end{document}